\documentclass[12pt]{elsarticle}
\usepackage{bbm}
\usepackage{amsfonts}
\usepackage{amsmath}
\usepackage{amssymb}
\usepackage{graphics}
\usepackage{epsfig}
\usepackage{color}
\usepackage{times}
\usepackage{graphicx}
\usepackage{diagbox}
\usepackage{xcolor}

\journal{Physics Letters A}

\begin{document}

\begin{frontmatter}



\title{Entanglement swapping via three-step quantum walk-like protocol}


\author[ahu]{Xiao-Man Li}
\author[ahu]{Ming Yang}
\ead{mingyang@ahu.edu.cn}
\author[InsdeTele,UdeL]{Nikola Paunkovi\'c}
\ead{npaunkovic@gmail.com}
\author[hnu]{Da-Chuang Li}
\ead{dachuangli@ustc.edu.cn}
\author[hnu,ahu]{Zhuo-Liang Cao}
\address[ahu]{School of Physics and Material Science, Anhui University, Hefei, 230601, China}
\address[InsdeTele]{Instituto de Telecomunica\c{c}\~{o}es, Av. Rovisco Pais 1049-001, Lisboa, Portugal}
\address[UdeL]{Departamento de Matem\'atica, Instituto Superior T\'ecnico, Universidade de Lisboa, Avenida Rovisco Pais 1049-001, Lisboa, Portugal}
\address[hnu]{School of Electronic and Information Engineering, Hefei Normal University, Hefei, 230601, China}

\begin{abstract}
We present an entanglement swapping process for unknown nonmaximally entangled photonic states, where the standard Bell-state measurement is replaced by a three-step quantum walk-like state discrimination process, i.e., the practically nontrivial coupling element of two photons is replaced by manipulating their trajectories, which will greatly enrich the dynamics of the coupling between photons in realizing quantum computation, and reduce the integration complexity of optical quantum processing. In addition, the output state can be maximally entangled, which allows for entanglement concentration as well.
\end{abstract}


\end{frontmatter}

\section{introduction}
In quantum communication and quantum computation, quantum entanglement finds many significant applications, including quantum teleportation \cite{quantum teleportation}, quantum superdense coding~\cite{quantum superdense coding}, quantum cryptography \cite{quantum cryptography}, etc.  Typically, only maximally entangled states (MES) can lead to the perfect implementation of the above protocols. But in real experiments, unavoidable decoherence of quantum system is a serious hindrance to the realization of quantum information processing and quantum computation. In general, it is inevitable that the degree of entanglement decreases with the channel length, leading to an effective non-maximally entangled state. Undoubtedly, the use of non-MES could lead to severe decrease in the efficiency and fidelity of a quantum communication protocol. Therefore, creation of a MES from non-MESs attracts considerable attention in the community. To circumvent this problem, Schmidt projection scheme and Procrustean scheme have been proposed \cite{Bennett}. Although entanglement swapping scheme was proposed for entangling two remote qubits without direct interaction between them \cite{swapping, entanglemen swapping experimentally}, it can be regarded as an entanglement concentration method too \cite{puriviaswapping}.

 In the standard entanglement swapping process, a Bell state measurement constitutes the main swapping mechanism \cite{swapping}. But, the realization of a Bell-state measurement (BSM) is not an easy task in experiment, so efforts have been made to design the entanglement swapping schemes without BSM. For instance, several implementation schemes of the entanglement swapping without BSM have been proposed both in cavity QED systems \cite{QED1,QED2} and in quantum dot systems \cite{dot}. Essentially, the entanglement swapping schemes with or without BSM both require the coupling interactions between two qubits at the intermediate location. The coupling interactions in the swapping scheme with BSM can lead to a full discrimination of the four Bell states, meanwhile the coupling interactions in the swapping scheme without BSM can only lead to a partial discrimination of the four Bell states.

Recently, it was shown that quantum walk (QW) \cite{QW} can be used to implement a generalized measurement, i.e. a positive operator value measure (POVM) \cite{QWPOVM}, and furthermore, a generalized measurement has been realized in discriminating non-orthogonal quantum states by executing a properly engineered QW \cite{QWPOVMxue}. But, in these advances, only the non-orthogonal quantum state discrimination of a single qubit has been studied and realized via QW. If this QW based state discrimination process can be generalized to the two-qubit case, it can be used to implement entanglement swapping too. In this paper, we present a three-step QW-like state discrimination scheme for four non-orthogonal two-qubit states, and thus the entanglement swapping scheme for two unknown non-maximally entangled states. The output state of the swapping process is maximally entangled, which allows for entanglement concentration as well.

In addition, the coupling between two qubits is the core part of the entanglement swapping schemes. But the current existing coupling mechanisms for photonic qubits and matter qubits are not suitable for integration. The coupling mechanism for two photons in our entanglement swapping process is realized by manipulating the trajectories of the photons. Thus, it is very easy to implement and integrate, and the versatile site-dependent operations and the intersite trajectory manipulations will greatly enrich the dynamics that this  process can produce. Since our protocol can formally be described as a two-particle three-step QW, this opens new possibilities for quantum computation using the existing optical implementations of QWs.

This paper is organized as follows. In Sec.~\ref{sec:ent_swap} we briefly introduce the concepts of entanglement swapping for non-maximally entangled states. In Sec.~\ref{sec:qw} we present our entanglement swapping scheme. Sec.~\ref{sec:conclusions} summarizes our results.

\section{Entanglement swapping for unknown non-maximally entangled states}
\label{sec:ent_swap}
Suppose there are two pairs of polarization-entangled photons (1, 2) and (3, 4) shared by three remote users Alice, Bob and Clare:
\begin{eqnarray}\label{1,2}
 |\psi \rangle _{12}&=&a \left|HH\right\rangle_{12} + b \left|VV\right\rangle_{12}, \\
  |\psi\rangle _{34}&=&a \left|HH\right\rangle_{34} + b \left|VV\right\rangle_{34},
 \end{eqnarray}
where $a$, $b$ satisfy the normalization condition $|a|^{2}+|b|^{2}=1$. Photons (1, 2) belong to Alice and Clare, respectively, and photons (3, 4) belong to Clare and Bob. Here, $|H\rangle$ ($|V\rangle$) denotes the horizontal (vertical) polarization state of the photons. Without loss of generality, we can assume that the superposition coefficients  $a$ and $b$ are all real numbers. Initially, the state of the two photon pairs is in a product form , which can be written as
\begin{eqnarray}\label{eq:1234}
 |\psi \rangle _{1234}&=&|\psi\rangle _{12}\otimes|\psi \rangle _{34} \nonumber\\
 &=&\sqrt{\frac{a^4+b^4}{2}}(|\psi\rangle _{14}^{+}|\psi\rangle _{23}^{1}+|\psi\rangle _{14}^{-}|\psi\rangle _{23}^{2}) \nonumber \\
 &&+ab(|\varphi\rangle _{14}^{+}|\psi\rangle_{23}^{3}+|\varphi\rangle _{14}^{-}|\psi\rangle _{23}^{4}),
 \end{eqnarray}
where
\begin{eqnarray}
\label{eq:psi}  | {{\psi }} \rangle _{14}^{\pm} & = & \frac{1}{\sqrt{2}}( \left|HH\right\rangle_{14} \pm  \left|VV\right\rangle)_{14}, \\
\label{eq:phi}  |\varphi \rangle _{14}^{\pm} & = & \frac{1}{\sqrt{2}}( \left|HV\right\rangle_{14} \pm  \left|VH\right\rangle_{14}), \\
\label{eq:1}  | {{\psi }} \rangle _{23}^{1} & = & \frac{1}{\sqrt{a^4+b^4}}(a^2 \left|HH\right\rangle_{23} + b^2 \left|VV\right\rangle_{23}), \\
\label{eq:2}  | {{\psi }} \rangle _{23}^{2} & = & \frac{1}{\sqrt{a^4+b^4}}(a^2 \left|HH\right\rangle_{23} -b^2  \left|VV\right\rangle_{23}), \\
\label{eq:3}  | {{\psi }} \rangle _{23}^{3} & = & \frac{1}{\sqrt{2}}( \left|HV\right\rangle_{23} +  \left|VH\right\rangle_{23}), \\
\label{eq:4}  | {{\psi }} \rangle _{23}^{4} & = & \frac{1}{\sqrt{2}}( \left|HV\right\rangle_{23} - \left|VH\right\rangle_{23}).
\end{eqnarray}

\begin{figure}[!htbp]
\includegraphics[width=\columnwidth]{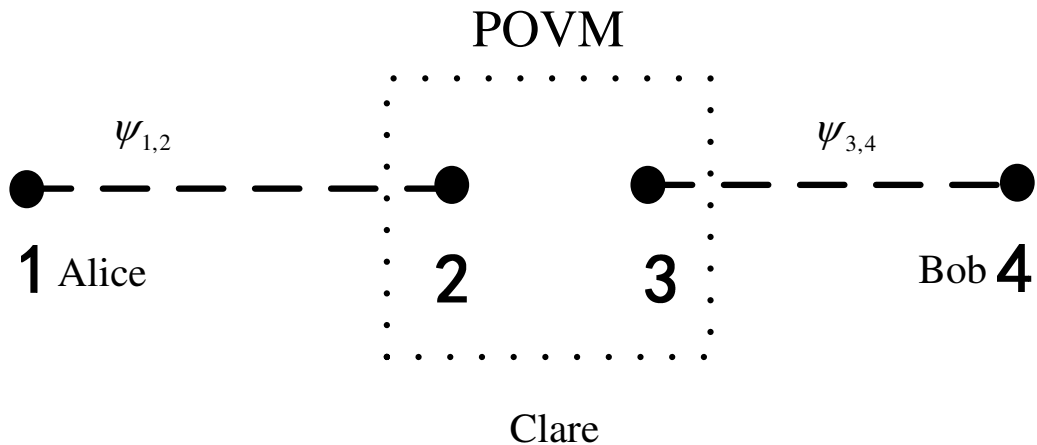}
\caption{The schematic diagram illustrating the procedure of entanglement swapping for unknown non-maximally entangled states.} \label{fig1}
\end{figure}

From Eq.~\eqref{eq:1234}, we can see that, as long as Clare, who has access to photons 2 and 3 (as depicted in Fig.~\ref{fig1}), can discriminate the four states in Eqs.~(\ref{eq:1}-\ref{eq:4}), four maximally entangled states in Eqs.~(\ref{eq:psi}, \ref{eq:phi}) can be generated among the two remote users Alice and Bob. But the four states in Eqs.~(\ref{eq:1}-\ref{eq:4}) are not orthogonal to each other, and they cannot be distinguished with unit probability. So generalized measurements (POVMs) must be introduced to discriminate these non-orthogonal states \cite{chefles,aldo}. Because the states to be swapped are unknown~\footnote{They are unknown in a sense that coefficients $a$ and $b$ are unknown, but the type of the states (superposition of both photon polarisations being either horizontal or vertical) is known.}, the states in Eqs.~(\ref{eq:1}, \ref{eq:2}) are unknown for us too, and thus these two states cannot be distinguished. Nevertheless, the states in Eqs.~(\ref{eq:3}, \ref{eq:4}) are totally known for us, so, in the following section, we will design a three-step QW-like process to discriminate these two states among the four non-orthogonal quantum states in Eqs.~(\ref{eq:1}-\ref{eq:4}).

 \section{Quantum walk-like swapping mechanism}
 \label{sec:qw}
 In this section, we are going to design a three-step QW-like scheme to distinguish the two states in Eqs.~(\ref{eq:3}, \ref{eq:4}) from the four non-orthogonal quantum states in Eqs.~(\ref{eq:1}-\ref{eq:4}), where the polarization degrees of photons 2, 3 are regarded as coin degrees of the three-step QW-like evolution, and the final position measurements on these two one-dimensional (1D) QW-like processes after three appropriately designed steps will tell us whether the discrimination succeeds or not.

Because our three-step scheme is a QW-like one, the state evolutions of the scheme are in similar forms as in QW systems, it is helpful for us to briefly review a standard 1D discrete-time QW \cite{QW}. The total Hilbert space of a walker consists of coin and position degrees of freedom, and is given by the tensor product $ H \equiv H_{\mathcal{C}}\otimes H_{\mathcal{P}}$ of two subspaces spanned by $\{ |H\rangle _{\mathcal{C}},|V\rangle _{\mathcal{C}} \}$ and $\{{|n\rangle _{\mathcal{P}},n\in \mathbb{Z}}\}$, respectively. Here, the subindex $\mathcal{C}$ denotes the coin degree, $\mathcal{P}$ denotes the position degree, and from now on they will be omitted for simplicity. One-step evolution of the system involves the coin flipping and conditional position shift based on the outcome of the coin flipping, and the corresponding unitary operation $U$ is
\begin{equation}
  U= S(C \otimes I),
\end{equation}
where $C\in U(2)$ is the coin flipping operator, $I$ is the identity operator in the position space, and the conditional position shift operator $S$ takes the form $S=\sum_{x}(|x+1\rangle\langle x|\otimes |H\rangle\langle H|+|x-1\rangle\langle x|\otimes |V\rangle\langle V|)$. Without loss of generality, we assume the walker is at the position $x=0$ initially, and the initial state of the coin is a superposition of $|H\rangle$ and $|V\rangle$ states. If the walk starts with the initial state $|\Psi(0)\rangle$, the final state of the system after $t$ steps becomes
\begin{equation}
 |\Psi(t)\rangle=U^t|\Psi(0)\rangle.
\end{equation}

In our scheme, the states to be distinguished are two-photon (photons 2 and 3 at Clare's location) joint states in Eqs.~(\ref{eq:1}-\ref{eq:4}) rather than the single-photon states, so the state evolutions for realizing this discrimination process are similar with the case of two walkers on two different lines. The joint Hilbert space of the two photons 2 and 3, on ``line 2'' and ``line 3'', respectively, is given by
\begin{equation}
 H_{23}\equiv H_{2}\otimes H_{3}\equiv (H_{\mathcal{C}2}\otimes H_{\mathcal{P}2})\otimes (H_{\mathcal{C}3}\otimes H_{\mathcal{P}3}).
\end{equation}
Here, $H_{2}$ and $H_{3}$ represent the Hilbert spaces of photons 2 and 3, respectively, each being isomorphic to the above Hilbert space $H$ defined for the one-particle case. Note that, since the two lines are different, the photons are fully distinguishable by their spatial positions, i.e., the Hilbert space labels 2 and 3 are the physical labels associated to line 2 and 3, respectively, and thus the effects of particle statistics are not present. We will therefore for simplicity omit the symmetrization postulate and keep working in the full Hilbert space $H$, rather than in its symmetric subspace. The case when the two photons' spatial states overlap, making them indistinguishable, will be discussed later in the text. The one-particle coin degree of freedom is given by a photon's polarization (thus the notation $|H\rangle$ and $|V\rangle$, for horizontal and vertical polarization along given axes $x$ and $y$, respectively). The corresponding unitary operation is given by
 \begin{equation}
  U_{23}= U_{2}\otimes U_{3},
\end{equation}
where both $U_{2}$ and $U_{3}$ are isomorphic to the $U$ defined above.

Assume both photons (``walkers'') are at the position $x=0$ initially, and the two polarization states (``coins'') are initially ($t=0$) prepared in a general (possibly entangled) state $|\Psi(0)\rangle_{\mathcal{C}_{23}}=\alpha|H\rangle_{\mathcal{C}2}|H\rangle_{\mathcal{C}3}+\beta|H\rangle_{\mathcal{C}2}|V\rangle_{\mathcal{C}3}+\gamma|V\rangle_{\mathcal{C}2}|H\rangle_{\mathcal{C}3}+\delta|V\rangle_{\mathcal{C}2}|V\rangle_{\mathcal{C}3}$ with the superposition coefficients satisfying the normalization condition. The overall composite system of the two photons is then in the following state (for simplicity, we write $|H,0\rangle_{2} $ to denote $|H\rangle_{\mathcal{C}2}|0\rangle_{\mathcal{P}2}$, etc.):
\begin{eqnarray}
 |\Psi(0)\rangle_{23} & = & \alpha|H,0\rangle_{2}|H,0\rangle_{3}+
 \beta|H,0\rangle_{2}|V,0\rangle_{3} \\
 && +\gamma|V,0\rangle_{2}|H,0\rangle_{3}+
 \delta|V,0\rangle_{2}|V,0\rangle_{3}. \nonumber
\end{eqnarray}

After $t$-step evolution, the final state of the system becomes $|\Psi(t)\rangle_{23} = U^t_{2}\otimes U^t_{3}|\Psi(0)\rangle_{23}$. For properly engineered evolution, the photons with different initial polarization states will arrive at different positions. By projective measurements on the photons' positions, the initial polarization states can be discriminated.

During the swapping process described by Eq.~\eqref{eq:1234}, the initial polarization state of Clare's two photons is one of the four states in Eqs.~(\ref{eq:1}-\ref{eq:4}), and the two photons both start from the origin position $x=0$. Thus, the four possible initial position-polarization (``walker-coin'') states are given by
 \begin{eqnarray}
  |\psi(0)\rangle _{23}^{1} \!\! & = & \!\! \frac{a^2|H,0\rangle_{2}|H,0\rangle_{3}+b^2|V,0\rangle_{2}|V,0\rangle_{3}}{\sqrt{a^4+b^4}}, \\
  |\psi(0)\rangle _{23}^{2} \!\! & = & \!\! \frac{a^2|H,0\rangle_{2}|H,0\rangle_{3}-b^2|V,0\rangle_{2}|V,0\rangle_{3}}{\sqrt{a^4+b^4}}, \\
  |\psi(0)\rangle _{23}^{3} \!\! & = & \!\! \frac{|H,0\rangle_{2}|V,0\rangle_{3} + |V,0\rangle_{2}|H,0\rangle_{3}}{\sqrt{2}}, \\
  |\psi(0)\rangle _{23}^{4} \!\! & = & \!\! \frac{|H,0\rangle_{2}|V,0\rangle_{3} - |V,0\rangle_{2}|H,0\rangle_{3}}{\sqrt{2}}.
\end{eqnarray}

\begin{figure}[!htbp]
\includegraphics[width=\columnwidth]{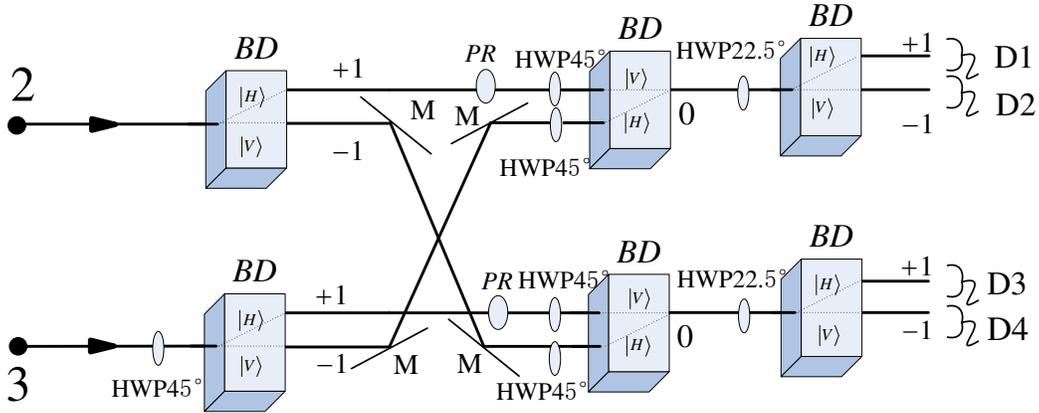}
\caption{Schematic drawing of the optical circuit for the three-step QW-like scheme that partially distinguishes the four possible two-photon polarization states. BD indicates beam displacers, by which vertically polarized photons are directly transmitted and horizontally polarized photons are moved up into a neighboring mode with a lateral displacement. M denotes mirrors, and PR indicates phase retarder, which is used for phase compensation. HWP$45^{\circ}$ is a half-wave plate oriented at $45^{\circ}$, whose function can be expressed as $|H\rangle \rightarrow|V\rangle$  and $|V\rangle \rightarrow |H\rangle$. HWP$22.5^{\circ}$ is a half-wave plate oriented at $22.5^{\circ}$, which can induce the transformations $|H\rangle \rightarrow  \frac{|H\rangle+|V\rangle}{\sqrt{2}}$ and $|V\rangle \rightarrow \frac{|H\rangle-|V\rangle}{\sqrt{2}}$. $D_i$ $(i=1,2,3,4)$ are ideal photon detectors.} \label{fig2}
\end{figure}

The detailed three-step swapping mechanism is shown in Fig. \ref{fig2}. The polarization operators for each photon depend on the step and, for the first step, on the line as well (the polarization operator for line 2 is in the first step the identity $I$, while for line 3 the polarization operator is the NOT gate, with respect to the $\{|H\rangle,|V\rangle \}$ basis). The polarization operators are labelled by $C_{i,j}$, where the label $i = 1,2,3$ represents the step, while the label $j = 2,3$ represents the line. Thus, in the first step the polarization operators are $C_{1,2} = I$ and
\begin{equation}
C_{1,3}=
\left(
\begin{array}{ccc}
 0 & 1  \\
 1 & 0
 \end{array}
\right).
\end{equation}
The polarization operators in the second step are also simple NOT gates,
\begin{equation}
C_{2,2} = C_{2,3} =
\left(
\begin{array}{ccc}
 0 & 1  \\
 1 & 0
 \end{array}
\right),
\label{C:22,23}
\end{equation}
while in the third step polarization operators are Hadamard gates,
\begin{equation}
C_{3,2} = C_{3,3} =
\frac{1}{\sqrt{2}}
\left(
\begin{array}{ccc}
 1 & 1  \\
 1 & -1
 \end{array}
\right).
\label{C:32,33}
\end{equation}
Each polarization rotation is followed by a conditional position shift operation $S$. The polarization operations can be implemented by half wave plates (HWPs) set at different orientations, and the conditional position shift operation $S$ can be realized by birefringent calcite beam displacers (BDs).

Hence, at the end of the first step, the states are
 \begin{eqnarray}
  |\psi(1)\rangle _{23}^{1} \!\! & = & \!\! \frac{a^2|H,1\rangle_{2}|V,-1\rangle_{3}+b^2|V,-1\rangle_{2}|H,1\rangle_{3}}{\sqrt{a^4+b^4}}, \\
  |\psi(1)\rangle _{23}^{2} \!\! & = & \!\! \frac{a^2|H,1\rangle_{2}|V,-1\rangle_{3}-b^2|V,-1\rangle_{2}|H,1\rangle_{3}}{\sqrt{a^4+b^4}}, \\
  |\psi(1)\rangle _{23}^{3} \!\! & = & \!\! \frac{|H,1\rangle_{2}|H,1\rangle_{3}+|V,-1\rangle_{2}|V,-1\rangle_{3}}{\sqrt{2}}, \\
  |\psi(1)\rangle _{23}^{4} \!\! & = & \!\! \frac{|H,1\rangle_{2}|H,1\rangle_{3}-|V,-1\rangle_{2}|V,-1\rangle_{3}}{\sqrt{2}}.
\end{eqnarray}

After the first step, the two site-dependent (x = -1) paths of the two photons will be exchanged, which is the key part of this swapping mechanism and causes the coupling between the two photons. Note that this kind of coupling is simply done by redirecting and exchanging the paths of two photons, which is much simpler than the coupling induced by a beam splitter. As immediately after the exchange the two photons are still spatially distinguishable (in case both photons end up in line 2, the one that came from line 2 is now at position +1 while the one coming from line 3 is at position -1; in case both are at line 3, the roles are symmetric -- the photon that stayed in line 3 is at position +1, while the other is at position -1), we can keep omitting symmetrization postulate. Nevertheless, as the position degrees of freedom of both photons are, after the exchange, enlarged, we will additionally label the position quantum numbers by $n_2$ and $n_3$ for the positions on line 2 and line 3, respectively, with $n_{2/3}\in \mathbb{Z}$. That is to say, the lines 2, 3 are exactly two different spatial locations, which enlarge the position space of the photon from the other line, for  instance, line 2 is an enlargement of the position space of the photon in line 3 and vice versa. In other words, the relevant part of the ``second photon'' Hilbert space (the photon coming from the line 2) is now $H_{2} = \mbox{span}\{ |-1_2\rangle _{2}, |0_2\rangle _{2}, |+1_2\rangle _{2}, |-1_3\rangle _{2}, |0_3\rangle _{2}, |+1_3\rangle _{2} \}$, and analogously for $H_3$. After the path exchange, the states evolve into
 \begin{eqnarray}
\!\!\!\!\!\! |\psi(1)\rangle _{23}^{1} \!\!\!\! & = & \!\!\!\! \frac{a^2|H,1_2\rangle_{2}|V,-1_2\rangle_{3}+b^2|V,-1_3\rangle_{2}|H,1_3\rangle_{3}}{\sqrt{a^4+b^4}}, \\
\!\!\!\!\!\!  |\psi(1)\rangle _{23}^{2} \!\!\!\! & = & \!\!\!\! \frac{a^2|H,1_2\rangle_{2}|V,-1_2\rangle_{3}-b^2|V,-1_3\rangle_{2}|H,1_3\rangle_{3}}{\sqrt{a^4+b^4}}, \\
\!\!\!\!\!\!  |\psi(1)\rangle _{23}^{3} \!\!\!\! & = & \!\!\!\! \frac{|H,1_2\rangle_{2}|H,1_3\rangle_{3}+|V,-1_3\rangle_{2}|V,-1_2\rangle_{3}}{\sqrt{2}}, \\
\!\!\!\!\!\!  |\psi(1)\rangle _{23}^{4} \!\!\!\! & = & \!\!\!\! \frac{|H,1_2\rangle_{2}|H,1_3\rangle_{3}-|V,-1_3\rangle_{2}|V,-1_2\rangle_{3}}{\sqrt{2}}.
\end{eqnarray}
Note that for the first two states, $|\psi(1)\rangle _{23}^{1/2}$, the two photons end in the same line, while for the other two, $|\psi(1)\rangle _{23}^{3/4}$, they end up in different lines.

For the second step the polarization operators are expressed in Eq. (\ref{C:22,23}). To make sure that when the two photons end up in the same line at the end of step 2 their spatial wave-functions fully overlap, we introduce two phase retarders (PRs), which are used for phase compensation.  As their spatial wave-functions fully overlap, being two identical particles, the two photons would be indistinguishable after step 2, and the effects of quantum statistics would occur, i.e. the first two states $|\psi(2)\rangle _{23}^{1/2}$ would consist only of the ``bunching" terms (a characteristic behaviour of indistinguishable bosons). Therefore, the four states after step 2 become (for simplicity, we now factor the overall polarization and position states; also, since the Hilbert space labels are now redundant, we omit them as well)
\begin{eqnarray}
\!\!  |\psi(2)\rangle^{1} \!\!\!\! & = & \!\!\!\! \frac{|H,V\rangle + |V,H\rangle}{\sqrt{2}}\otimes\frac{a^2|0_2,0_2\rangle + b^2|0_3,0_3\rangle}{\sqrt{a^4+b^4}}, \\
%
\!\!  |\psi(2)\rangle^{2} \!\!\!\! & = & \!\!\!\! \frac{|H,V\rangle + |V,H\rangle}{\sqrt{2}}\otimes\frac{a^2|0_2,0_2\rangle - b^2|0_3,0_3\rangle}{\sqrt{a^4+b^4}}, \\
%
\!\!  |\psi(2)\rangle^{3} \!\!\!\! & = & \!\!\!\! \frac{|V,V\rangle + |H,H\rangle}{\sqrt{2}}\otimes |0_2,0_3\rangle, \\
\!\!  |\psi(2)\rangle^{4} \!\!\!\! & = & \!\!\!\! \frac{|V,V\rangle - |H,H\rangle}{\sqrt{2}}\otimes |0_2,0_3\rangle.
\end{eqnarray}


For the third step the polarization operators are expressed in Eq. (\ref{C:32,33}). After the three-step evolution, the initial four possible position-polarization (``walker-coin'') states become
\begin{eqnarray}
\label{eq:31i}	
|\psi (3) \rangle^{1} \!\!\!\! & = & \!\!\!\! \frac{1}{\sqrt{2}}  \Big[  |H,H\rangle\otimes \Big(\frac{a^2|1_2,1_2\rangle + b^2|1_3,1_3\rangle}{\sqrt{a^4+b^4}}\Big) \\ & & - |V,V\rangle\otimes \Big(\frac{a^2|-1_2,-1_2\rangle + b^2|-1_3,-1_3\rangle}{\sqrt{a^4+b^4}}\Big)\Big], \nonumber \\
\label{eq:32i}
  |\psi (3) \rangle^{2} \!\!\!\! & = & \!\!\!\! \frac{1}{\sqrt{2}}  \Big[  |H,H\rangle\otimes \Big(\frac{a^2|1_2,1_2\rangle - b^2|1_3,1_3\rangle}{\sqrt{a^4+b^4}}\Big) \\ & & - |V,V\rangle\otimes \Big(\frac{a^2|-1_2,-1_2\rangle - b^2|-1_3,-1_3\rangle}{\sqrt{a^4+b^4}}\Big)\Big], \nonumber \\
\label{eq:33i}
  |\psi(3)\rangle^{3} \!\!\!\! & = & \!\!\!\! \frac{1}{\sqrt{2}}\Big(|H,H\rangle \! \otimes \! |1_2,1_3\rangle \! + \! |V,V\rangle \! \otimes \! |-1_2,-1_3\rangle \Big), \\
\label{eq:34i}
  |\psi(3)\rangle^{4} \!\!\!\! & = & \!\!\!\! \frac{-1}{\sqrt{2}}\Big(|H,V\rangle \! \otimes \! |1_2,-1_3\rangle \! + \! |V,H\rangle\! \otimes \! |-1_2,1_3\rangle \Big).
\end{eqnarray}

At the end of the third step, four photon detectors will be used to detect the position information of the two photons. From Eqs.~(\ref{eq:31i}-\ref{eq:34i}), we can see that the position information of the two photons will help us to identify two polarization states in Eqs.~(\ref{eq:3},\ref{eq:4}), i.e., the ``successful" results are coming from the cases when each line ends up containing only one photon, while the ``unsuccessful" ones are characterised by the fact that both photons end up in the same line. Specifically, if we observe a coincidence either between detectors $D1$ and $D3$ or $D2$ and $D4$, then the incident two-photon polarization state is $|\psi\rangle_{23}^{3}$. On the other hand, if we observe a coincidence between detectors $D1$ and $D4$ or $D2$ and $D3$, then the incident two-photon polarization state is $|\psi\rangle_{23}^{4}$. The other two incident polarization states will both lead only to single-detector clicks, such that either of the four detectors $D1-D4$ can click for both initial states $|\psi\rangle_{23}^{1}$ and $|\psi\rangle_{23}^{2}$, which correspond to an inconclusive result. The correspondence between the results of the coincidence measurements and the incident two-photon states is listed in Table~\ref{tab1}.

\begin{table}[ht]
\caption{The results of the coincidence measurements versus the incident two-photon polarization states.}
\centering{\begin{tabular}{p{4.5cm} p{4.5cm}} \hline
            Initial polarization states                             &  Clicks \\   \hline
       $ |\psi\rangle _{23}^{1}$                 & $D_{1}$, or $D_{2}$, or $D_{3}$, or $D_{4}$ \\
       $ |\psi\rangle _{23}^{2}$                 & $D_{1}$, or $D_{2}$, or $D_{3}$, or $D_{4}$ \\
       $ |\psi\rangle _{23}^{3}$                 & $D_{1}$  and $D_{3}$, or $D_{2}$  and $D_{4}$ \\
       $ |\psi\rangle _{23}^{4}$                 & $D_{1}$  and $D_{4}$, or $D_{2}$  and $D_{3}$ \\
       \hline
       \end{tabular}}
       \label{tab1}
       \end{table}

After the identification of the state $|\psi\rangle _{23}^{3}$ or $|\psi\rangle _{23}^{4}$, the state of the two remote photons 1 and 4 is projected onto the maximally entangled state $|\varphi\rangle _{14}^{+}$ or $|\varphi\rangle _{14}^{-}$. That is to say, the entanglement swapping for unknown non-maximally entangled states is realized by two three-step QW-like operations, and thus the unknown non-maximally entangled states are concentrated into maximally entangled ones.

One may wonder if the swapping process might succeed when the spatial wave-functions of the two photons do not overlap at the end of step 2 (in general, this happens). Does the problem of temporal synchronicity of the two photons affect the final results? The following analysis shows that the problem of temporal synchronicity of the two photons is irrelevant to our swapping scheme. If the two photons from line 2 and line 3 who ``meet" during the second shift operation have different times of arrival to the third BD, they would be distinguishable, and the results presented in (\ref{eq:31i}) and (\ref{eq:32i}) would be (note that now the Hilbert space labels 2 and 3 have physical meaning, denoting the different times of arrival of the two photons; also, for simplicity we omit the overall normalization $2\sqrt{a^4+b^4}$)
\begin{eqnarray}
\label{eq:31d}
\!\!\!\!\!\!\!\! |\psi^\prime(3)\rangle _{23}^{1} & \!\! \propto \!\! & a^2\Big( |H,1_2\rangle_2|H,1_2\rangle_3 + |H,1_2\rangle_2|V,-1_2\rangle_3 \nonumber  \\ & \!\! - \!\! & |V,-1_2\rangle_2|H,1_2\rangle_3- |V,-1_2\rangle_2|V,-1_2\rangle_3 \Big )                                                    \nonumber  \\
 & \!\! + \!\! & b^2\Big( |H,1_3\rangle_2|H,1_3\rangle_3 - |H,1_3\rangle_2|V,-1_3\rangle_3 \nonumber  \\ & \!\! + \!\! & |V,-1_3\rangle_2|H,1_3\rangle_3- |V,-1_3\rangle_2|V,-1_3\rangle_3 \Big ), \\
\label{eq:32d}
\!\!\!\!\!\!\!\!  |\psi^\prime(3)\rangle _{23}^{2} & \!\! \propto \!\! & a^2\Big( |H,1_2\rangle_2|H,1_2\rangle_3 + |H,1_2\rangle_2|V,-1_2\rangle_3 \nonumber  \\ & \!\! - \!\! & |V,-1_2\rangle_2|H,1_2\rangle_3- |V,-1_2\rangle_2|V,-1_2\rangle_3 \Big )                                                    \nonumber  \\
 & \!\! - \!\! & b^2\Big( |H,1_3\rangle_2|H,1_3\rangle_3 - |H,1_3\rangle_2|V,-1_3\rangle_3 \nonumber  \\ & \!\! + \!\! & |V,-1_3\rangle_2|H,1_3\rangle_3- |V,-1_3\rangle_2|V,-1_3\rangle_3 \Big ),
\end{eqnarray}
and the resulting states in Eqs.~(\ref{eq:33i},\ref{eq:34i}) do not change. From the Eqs.~(\ref{eq:33i}-\ref{eq:32d}), we can see that the state $|\psi\rangle _{23}^{3}$ or $|\psi\rangle _{23}^{4}$ still can be identified by the coincidence measurements listed in Table \ref{tab1}. That is to say, the fact that one detector from each line clicks, as opposed to the case when ones from only one line click, would discriminate the ``successful" swapping results from the ``unsuccessful" ones. So the temporal asynchronization of the two photons does not alter the final result of the paper. Naturally, one may prefer the asynchronous case, because one can always delay one photon over the other, making the two photons fully distinguishable.

From the discussions in Ref. \cite{QWPOVMxue,broome10,xue15}, we can see that our scheme is feasible and can be implemented experimentally. The two photons $2, 3$  can be injected into the two free-space modes $0$ of two BDs. Subsequently they propagate through BDs, free-space modes $\pm1$ between BDs,  and recombine at the second set of BDs, which forms an interferometric structure. After the second  interferometric structure, the output photons will be coupled into single-mode fibers and subsequently detected by photon detectors (avalanche photodiodes). It was demonstrated that the interferometric structure used here is inherently stable, and no active phase locking is required \cite{broome10},  which demonstrates the feasibility of the current scheme too.

From the results of the scheme we can see that the `click' or `no-click' indicating the presence or the absence of photons in the corresponding mode is sufficient for us to confirm the success of the scheme, which means that only common single-photon detectors will be involved rather than the sophisticated photon-number-resolution detectors \cite{wildfeuer,miller,lita}. In addition, only two photons are involved in the swapping dynamics, so the two-photon coincidence measurements indicate and confirm the success of the swapping process. One may say that, in a real situation, a successful output may not induce a two-photon coincidence counting because of the non-perfect detection efficiency of photon detectors. Yes, it does happen sometimes, but this fact only decreases the success probability of the scheme without decreasing the fidelity of the output state. As long as a two-photon coincidence counting is registered, the swapping process succeeds. Currently, the detection efficiency of a photon detector can reach $93\%$ \cite{SPD}, which only slightly decreases the success probability $2|ab|^{2}$ (see~\eqref{eq:1234}) of our swapping scheme. On the other hand, the system error, caused by the imperfections of the optical components such as the dark counts of photon detectors, the nonplanar optical surfaces and the inaccurate angles of BDs and wave plates, will cause a imperfect fidelity $0.99$ of the output state in our scheme \cite{QWPOVMxue,broome10,xue15}.

\section{Conclusion}
\label{sec:conclusions}
In this paper, we demonstrated that much richer dynamics for two-photon states can be produced via the quantum walk-like evolutions, such as  site-dependent operations and the intersite trajectory manipulations, which opens a new direction in realizing quantum computation. We proposed a three-step quantum walk-like entanglement swapping scheme  for two unknown non-maximally entangled states. The swapping mechanism used here is realized by manipulating the trajectories of two photons, which greatly reduces the integration complexity. In addition, maximally entangled states are generated among two spatially separated particles from two unknown non-maximally entangled states, which means an entanglement concentration process can be realized via our scheme too.

The result presented in this paper is just a demonstration of a simple QW-like dynamics for two-photon states, and more new dynamics for quantum communication and quantum computation, such as the basic two-photon logic gates, will be our next study direction.

\section*{Acknowledgments}
This work is supported by the National Natural Science Foundation of China (NSFC) under Grants No.11274010, No.11204061, No.11374085; Anhui Provincial Natural Science Foundation under Grant No. 1408085MA16; The Anhui Provincial Candidates for academic and technical leaders Foundation under Grant No. 2015H052. NP acknowledges the support of SQIG Security and Quantum Information Group and IT project QbigD funded by FCT PEst-OE/EEI/LA0008/2013 and UID/EEA/50008/2013.

\end{document}